\documentclass[]{spie}  

\usepackage{amsmath,amsfonts,amssymb}
\usepackage{graphicx}
\usepackage[colorlinks=true, allcolors=blue]{hyperref}

\newcommand\blfootnote[1]{%
	\begingroup
	\renewcommand\thefootnote{}\footnote{#1}%
	\addtocounter{footnote}{-1}%
	\endgroup
}

\title{SMART tracking: Simultaneous anatomical imaging and real-time passive device tracking for MR-guided interventions}

\author[a]{Frank Zijlstra}
\author[a]{Max A. Viergever}
\author[a]{Peter R. Seevinck}

\affil[a]{Image Sciences Institute, University Medical Center Utrecht, Utrecht, The Netherlands}

\begin{document}
\maketitle

\blfootnote{This work has been published in Physica Medica: European Journal of Medical Physics, 2019, Volume 64, 252 - 260, DOI: \url{https://doi.org/10.1016/j.ejmp.2019.07.019}.

\textcopyright \space 2019. This manuscript version is made available under the CC-BY-NC-ND 4.0 license \url{http://creativecommons.org/licenses/by-nc-nd/4.0/}}

\begin{abstract}
\textbf{Purpose:} This study demonstrates a proof of concept of a method for simultaneous anatomical imaging and real-time (SMART) passive device tracking for MR-guided interventions.

\noindent\textbf{Methods:} Phase Correlation template matching was combined with a fast undersampled radial multi-echo acquisition using the white marker phenomenon after the first echo. In this way, the first echo provides anatomical contrast, whereas the other echoes provide white marker contrast to allow accurate device localization using fast simulations and template matching. This approach was tested on tracking of five 0.5 mm steel markers in an agarose phantom and on insertion of an MRI-compatible 20 Gauge titanium needle in ex vivo porcine tissue. The locations of the steel markers were quantitatively compared to the marker locations as found on a CT scan of the same phantom.

\noindent\textbf{Results:} The average pairwise error between the MRI and CT locations was 0.30 mm for tracking of stationary steel spheres and 0.29 mm during motion. Qualitative evaluation of the tracking of needle insertions showed that tracked positions were stable throughout needle insertion and retraction. 

\noindent\textbf{Conclusions:} The proposed SMART tracking method provided accurate passive tracking of devices at high framerates, inclusion of real-time anatomical scanning, and the capability of automatic slice positioning. Furthermore, the method does not require specialized hardware and could therefore be applied to track any rigid metal device that causes appreciable magnetic field distortions.
\end{abstract}


\section{Introduction}
MR-guided interventions have shown promise in a variety of applications, including needle biopsies \cite{fischbach_mr-guided_2012,krafft_passive_2013}, vascular interventions \cite{bakker_mr-guided_1997,draper_passive_2006,seppenwoolde_fully_2006,karimi_position_2016}, and MR-guided radiation therapy \cite{kapur_3-t_2012,wang_real-time_2015}. An important challenge for MR-guided interventions is fast and accurate localization of interventional devices. Most interventional devices used in MRI, such as metal needles and paramagnetic markers, do not generate contrast at the exact location of the devices. Instead, the presence of these devices causes artifacts in MR images due to magnetic susceptibility differences. The shape of these artifacts is non-trivial and can interfere with accurate localization of the devices \cite{lagerburg_simulation_2008}. For example, the signal void caused by a device is not necessarily representative for the actual location of the device, because the shape of the void can change depending on the acquisition parameters and the device orientation \cite{lagerburg_simulation_2008,wachowicz_characterization_2006}, and because the device void can be confounded with nearby anatomical signal voids.

Current techniques for tracking devices in MRI can be broadly classified into passive, semi-active, and active tracking techniques \cite{duerk_brief_2001}. In passive tracking, the device is localized based on its passive effect on the MR signal. The artifacts caused by magnetic field changes induced by the presence of a metal device can be detectable, either in anatomical images from standard pulse sequences \cite{kapur_3-t_2012,lagerburg_simulation_2008,wachowicz_characterization_2006} or in dedicated pulse sequences \cite{draper_passive_2006,karimi_position_2016,zhang_tracking_2013}. Alternatively, markers filled with contrast fluid can be added to devices to make them detectable in MR images \cite{krafft_passive_2013,larson_optimized_2012}.
 
Passive tracking methods can either generate positive contrast which can be visualized on an anatomical reference \cite{draper_passive_2006,patil_automatic_2009}, or use passive signal effects to determine the exact location and orientation of the device \cite{krafft_passive_2013,karimi_position_2016,zhang_tracking_2013,larson_optimized_2012}. Passive tracking has shown promise in a variety of applications \cite{fischbach_mr-guided_2012,bakker_mr-guided_1997,seppenwoolde_fully_2006}. The accuracy and framerate achieved by passive tracking are mostly limited by the strength of the passive effect of the device, i.e. larger devices and devices with strong magnetic susceptibilities will be easier to track. Dedicated pulse sequences that generate more specific contrast around the device can also enable faster, more accurate tracking \cite{karimi_position_2016,zhang_tracking_2013}.

(Semi-)active tracking methods use specialized hardware to overcome the limitations of passive tracking and provide fast and accurate tracking. In both active and semi-active tracking, small RF coils are attached to interventional devices \cite{duerk_brief_2001}. In the case of active tracking, these coils are attached to a receive channel on the scanner. The signal generated by these coils is very specific for the location of the device, which can be localized by acquiring only a few one-dimensional projections \cite{wech_measurement_2014}. This process is fast, and can in principle be interleaved with regular real-time scanning protocols to provide an anatomical reference for the localized device \cite{buecker_simultaneous_2002,saikus_mri-guided_2011}. The biggest disadvantage of (semi-)active tracking is that specialized hardware is required, which is costly to develop and adds to the size of the devices. Furthermore, the hardware causes additional RF safety concerns due to potential heating \cite{konings_heating_2000}.

We believe that in an ideal situation an MR-based device tracking method should share the advantages of both passive and active tracking, while minimizing the disadvantages. First, this means that the method must be accurate, robust, and should have real-time updates for device tracking (i.e. multiple updates per second). Second, the system should allow exact visualization of the device on an anatomical reference image, of which the slice position should automatically update. Ideally, this image would be acquired simultaneously to ensure that patient motion and deformation of anatomical structures does not influence the accuracy of the visualization. Finally, the hardware used in the method should be safe, cheap to implement, and flexible with regard to clinical applications.

In this study, we developed a passive tracking method which aims to satisfy these criteria. We propose SMART tracking: SiMultaneous Anatomical imaging and Real-Time tracking. This method builds on previous research on dephased MRI \cite{bakker_dephased_2006} and the white marker phenomenon \cite{seppenwoolde_passive_2003} to provide selective positive contrast near metal devices, and fast simulation and Phase Correlation template matching \cite{krafft_passive_2013,zijlstra_fast_2017} to exactly localize devices. An undersampled 2D radial multi-echo pulse sequence was used to achieve high update rates and to acquire anatomical contrast simultaneously with the device tracking. The proposed method requires no specialized hardware and can be applied to any metal device that induces sufficient magnetic field changes to locally cause dephasing. We demonstrate a proof of concept of the method on tracking of 0.5 mm steel markers in an agarose phantom and on insertion of an MRI-compatible 20 Gauge titanium needle in ex vivo porcine tissue.

The main innovations of this study with respect to previously published studies on metal device localization are the following: 1) Acceleration to real-time framerates through radial undersampling; 2) generalization of the Phase Correlation template matching and simulation methods to acquisitions that use non-Cartesian sampling, undersampling, and/or acquire multiple echoes; and 3) combination of anatomical contrast with positive contrast mechanisms to provide intrinsically registered anatomical reference for device localization.

\section{Methods}

The SMART tracking method we propose in this study is a combination and extension of multiple previously proposed methodologies for positive contrast and device localization. The acquisition is an undersampled radial 4-echo acquisition in which the white marker phenomenon \cite{seppenwoolde_passive_2003} was induced in the 2nd to 4th echoes. The white marker phenomenon is a positive contrast mechanism that uses a dephasing gradient to counteract intravoxel dephasing caused by the presence of metal device, while background signal is dephased. Device localization was performed using Phase Correlation template matching \cite{zijlstra_fast_2017}, for which the templates were simulated with the FORECAST method \cite{zijlstra_fast_2017}. We adapted the simulation to include non-Cartesian sequences and the white marker phenomenon. Finally, the device was tracked using a Kalman filter.

In the following sections we describe the separate components of the SMART tracking method in more detail. Finally we describe the experiments we performed on phantoms with two different types of metal devices: small stainless steel spherical markers and an MRI-compatible needle.

\subsection{Acquisition}

The basic pulse sequence used in this study was an undersampled radial 2D gradient echo scan with four echoes (resolution = $1.2 \times 1.2 \times 15$ mm; field of view = $230 \times 230 \times 15$ mm; TE$_1$/TE$_2$/TE$_3$/TE$_4$/TR = 1.73/3.18/4.62/6.07/8.46 ms; flip angle = $10^\circ$; nr. of radial profiles = 192; dynamic scan time = 1.6 s), acquired at a field strength of 1.5T (Philips Achieva, Best, The Netherlands) using a 2-channel surface coil. We used a bit-reversed profile order \cite{chan_influence_2012} with an acceleration factor of 16, giving a scan time of 0.1 seconds per undersampled frame. This profile ordering allows approximately uniformly undersampled reconstructions up to an acceleration factor of 16.

\begin{figure}
	\centering
	
	\includegraphics[width=8cm]{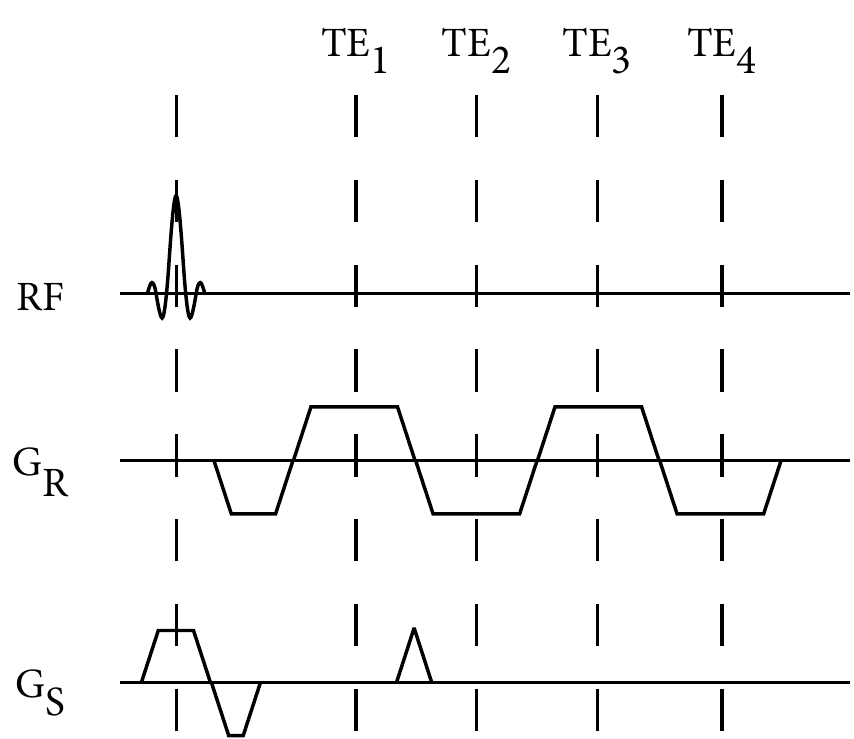}

	\caption{White marker pulse sequence for a gradient echo acquisition with four echoes. An additional dephasing gradient in the slice direction ($G_S$) was added between the first ($\mathit{TE}_1$) and second echo ($\mathit{TE}_2$) to induce the white marker phenomenon in the 2nd to 4th echoes.}
	\label{fig1}
\end{figure}

A dephasing gradient was added between the 1st and 2nd echo of the 4-echo acquisition to induce white marker contrast in the 2nd to 4th echoes, while the first echo yields a $T_1$-weighted anatomical image with minimal off-resonance artifacts. The strength of the gradient was chosen such that one complete cycle of dephasing (i.e. $2\pi$ radians) was induced over the slice thickness. The pulse sequence for this acquisition is shown in Figure \ref{fig1}.

\subsection{Reconstruction}

The first echo, with anatomical contrast, was reconstructed using a sliding window approach with a window of 16 frames, which yields a fully sampled k-space with a temporal resolution of 1.6 seconds. Prior to reconstruction, sampling density correction was applied to the radial profiles, followed by gridding using the NUFFT library \cite{fessler_nonuniform_2003}. Images from the two receive coils were combined using the sum of squares method.

For device localization, Phase Correlation (PC) images were reconstructed from single undersampled frames for all echoes. Because the PC method operates element-wise in the frequency domain, it can be applied to raw, undersampled non-Cartesian k-space data before image reconstruction, as long as the template data is also represented in the same non-Cartesian space. The reconstructed PC images were multiplied pixel-wise to yield one combined PC image. Implicitly, this means that in order to detect a device, the PC images for every echo must have a strong correlation peak at the device location.

The templates used in the PC matching were simulations of the metal device in a uniform background, using the scan parameters, undersampling scheme, and echo times specific for each echo and each frame. These simulations were performed with the FORECAST method \cite{zijlstra_fast_2017}, which we extended to allow simulation of the white marker phenomenon and non-Cartesian sampling trajectories, as described in Appendix A.

The devices were simulated at an isotropic resolution of 0.15 mm in a $25 \times 25 \times 15$ mm field of view (i.e. 6400 spins per pixel). The simulation models were based on physical models of the devices: The stainless steel spheres had a diameter of 0.5 mm and had an estimated magnetic susceptibility ($\chi$) of 5000 ppm. The needle was of type MRI Chiba (SOMATEX\textsuperscript{\textregistered}, Berlin, Germany), which consists of a titanium needle ($\chi$ = 190 ppm) and a nitinol mandrin (est. $\chi$ = 600 ppm). Only the artifacts around the needle tip were simulated, and the orientation of the needle was assumed to be known a priori.

Although in this study we applied the reconstruction retrospectively, the reconstruction was implemented such that it could be directly applied prospectively on k-space lines that are received sequentially during acquisition. We implemented the reconstruction pipeline in Matlab (Mathworks, Natick, Massachusetts, USA) and published the code on GitHub: \url{https://github.com/FrankZijlstra/SmartTracking}. Reconstruction time was around 0.1 second per frame on average, i.e. capable of reconstructing data in real-time.

\subsection{Tracking}

Although it is possible to extract positions from the PC image for each undersampled frame individually, we opted to use a tracking algorithm that incorporates knowledge of previous device locations to increase robustness. We implemented a Kalman filter \cite{kalman_new_1960} that tracks the position and velocity of the device, given position measurements from the PC image over time. The tracking was initialized by performing a fully sampled reconstruction of the first fully sampled frame of the acquisition. The initial device locations were found by locating the $N$ most intense local maxima in the PC image, where $N$ is the number of devices being tracked.

\subsection{Experimental setup}

We performed experiments with two different types of metal devices: small spherical steel markers (diameter 0.5 mm) and an MRI-compatible 20 Gauge biopsy needle. The primary goal of the experiments with the steel markers was to show the feasibility and to establish the accuracy of the proposed method. The steel markers can also be seen as a surrogate for other small markers that create dipolar field distortions, such as markers on a guidewire. The experiments with the needle serve as a proof of concept for clinical applications. The larger, orientation-dependent artifacts created by a needle are similar to other types of interventional devices, such as DBS-applicators and HDR brachytherapy sources.

For the first series of experiments we created a cylindrical phantom containing 5 steel markers in a cross pattern in a 2\% agarose gel. The phantom was scanned with our proposed tracking sequence in four different conditions: 1) stationary, 2) moving linearly along $B_0$ with varying speeds, 3) rotating in the coronal plane with varying speeds, and 4) rapidly moving and rotating at the same time.
 
For the second series of experiments we inserted an MRI Chiba (SOMATEX\textsuperscript{\textregistered}, Berlin, Germany) needle into ex vivo porcine tissue. The inserted needle was scanned with our proposed tracking sequence in three different conditions: 1) stationary, 2) needle inserted and retracted linearly along $B_0$ with varying speeds, and 3) needle inserted and retracted linearly at an approximately 45 degree angle with $B_0$ with varying speeds.

For the stationary experiments we also scanned the phantoms with anatomical contrast in all echoes and white marker contrast in all echoes. This allowed a comparison between anatomical and white marker contrast in both fully sampled and undersampled scans.

To validate the steel marker positions we acquired a CT scan of the marker phantom (resolution $0.21 \times 0.21 \times 0.67$ mm). The marker positions in the CT scan were located by finding the center of mass of connected components with voxel values larger than 2000 HU. We performed a rigid registration of the CT scan to a 3D gradient echo scan of the marker phantom (resolution $1 \times 1 \times 2$ mm), which was acquired in the same session as the stationary experiment to serve as a reference. The CT marker positions were registered to the MRI coordinate space using the resulting transformation. In the stationary experiment, the tracked marker positions on MRI could be directly compared to the registered CT marker positions.

In the dynamic tracking experiment with the steel marker phantom, no direct comparison could be made to the CT marker positions, since the position of the phantom over time was unknown. Instead, for each frame we performed a rigid registration of the CT marker positions to the tracked positions in MRI using the Coherent Point Drift method \cite{myronenko_point_2010}. Any errors in the positions indicate deformation of the tracked configuration of the five markers with respect to the configuration found on CT.

Finally, we performed an experiment using dual plane tracking of a needle insertion. For this approach, 16-fold undersampled 2D acquisitions for coronal and sagittal slices were interleaved. The basic acquisition and reconstruction strategy remained unchanged. The tracking model was extended to 3D positions, where coronal slices updated the left-right and feet-head positions, and sagittal slices updated the anterior-posterior and feet-head positions. This means the left-right and anterior-posterior positions were updated at a rate of 5 Hz, while the feet-head position was updated at a rate of 10 Hz. Both the coronal and sagittal anatomical images fully updated once every 3.2 seconds.

\section{Results}

\subsection{Stationary devices}

\begin{figure}
	\centering
	
	\includegraphics[width=\textwidth]{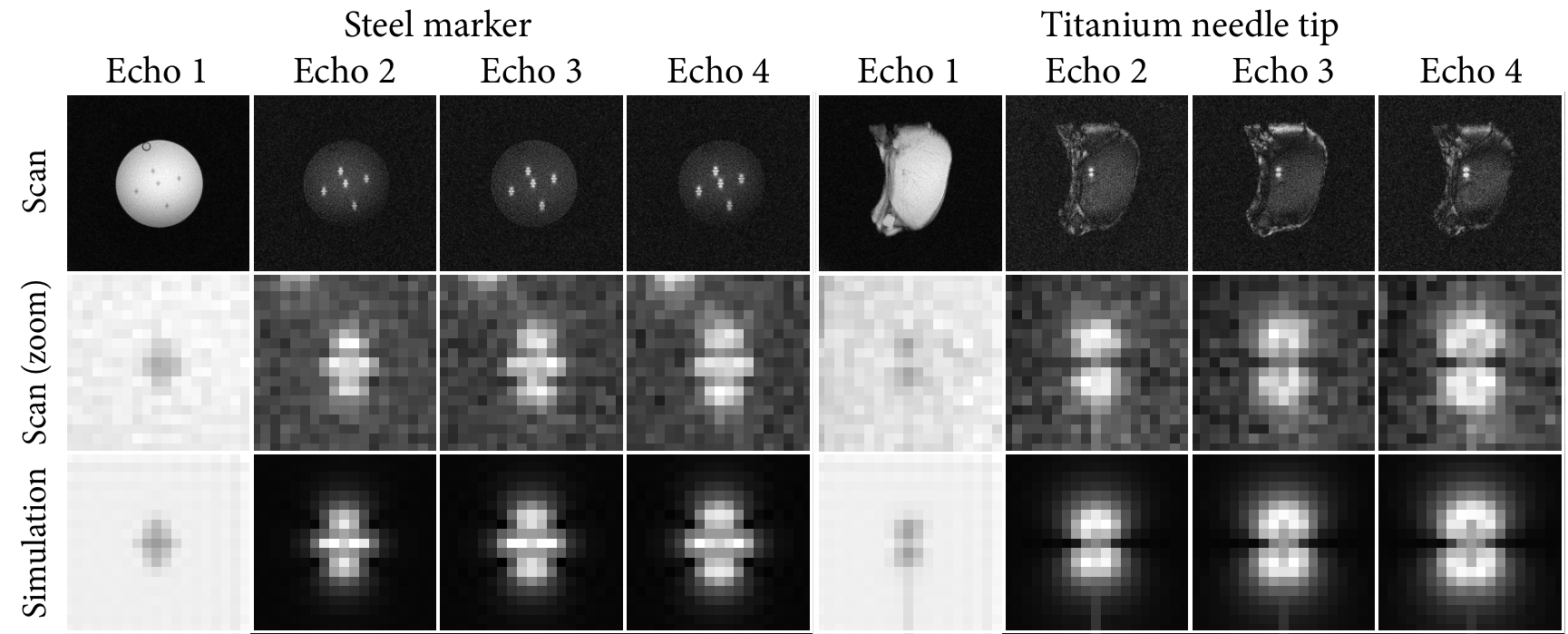}

	\caption{MRI scan and MRI simulation using the proposed multi-echo pulse sequence for a single steel marker (left) and tip of a titanium needle (right). The first echo shows anatomical contrast, whereas the 2nd to 4th echoes show white marker contrast.}
	\label{fig2}
\end{figure}

Figure \ref{fig2} shows the MRI scan and corresponding MRI simulation for a single steel marker and the tip of a titanium needle in our multi-echo pulse sequence with simultaneous anatomical and white marker contrast. Overall the shape and intensity of the simulated artifacts around the devices had a good correspondence with the actual MRI scan. Some small differences in the artifacts can be observed, which could be attributed to noise, partial volume effects due to sub-voxel shifts, and factors that were not included in the simulations, such as accurate simulation of the RF excitation pulse.

\begin{figure}
	\centering
	
	\includegraphics[width=\textwidth]{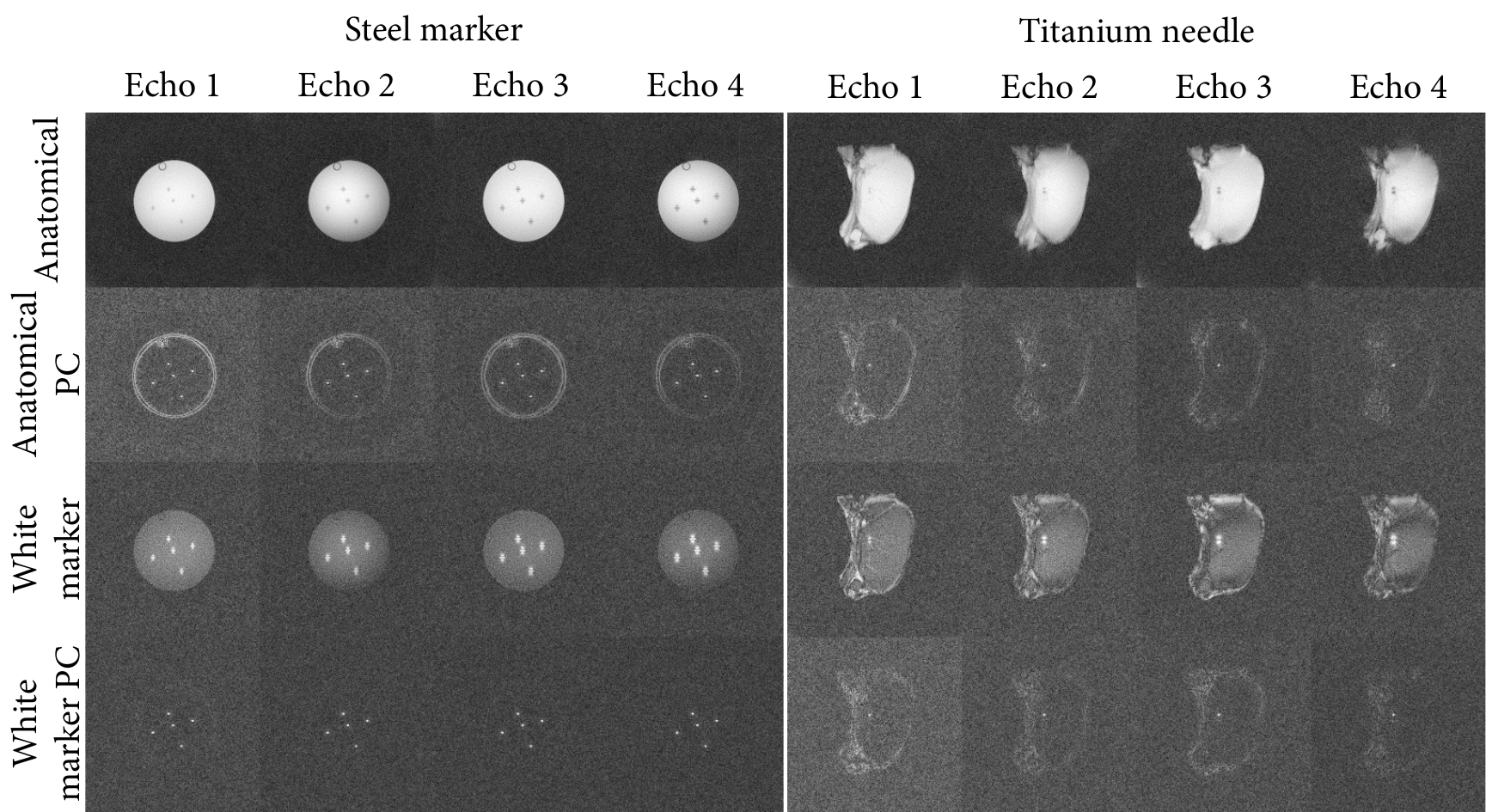}

	\caption{MRI scans and Phase Correlation (PC) maps on anatomical (rows 1 and 2) and white marker (rows 3 and 4) contrasts for the steel markers (left) and a titanium needle (right). Images are shown for all four echoes of the pulse sequences.}
	\label{fig3}
\end{figure}

Figure \ref{fig3} shows the difference between performing PC template matching on anatomical contrast and white marker contrast for the four echoes. First, as may be expected, the signal void related to intravoxel dephasing increases with increasing echo time. The same holds for the white marker phenomenon, the size of which increases with increasing echo time. For the white marker contrast we observed fewer high intensity correlations in anatomical structures than in the anatomical images, especially around sharp edges. In general we observed that the background dephasing in the white marker images was incomplete in some areas. This can be partially explained by structural variation over the slice, as well as presence of fatty tissue and bone. Additionally, inhomogeneity of the magnetic field appeared to have caused part of the incomplete dephasing. However, the incomplete dephasing did not appear to have a major influence on the PC images.

\begin{figure}
	\centering
	
	\includegraphics[width=\textwidth]{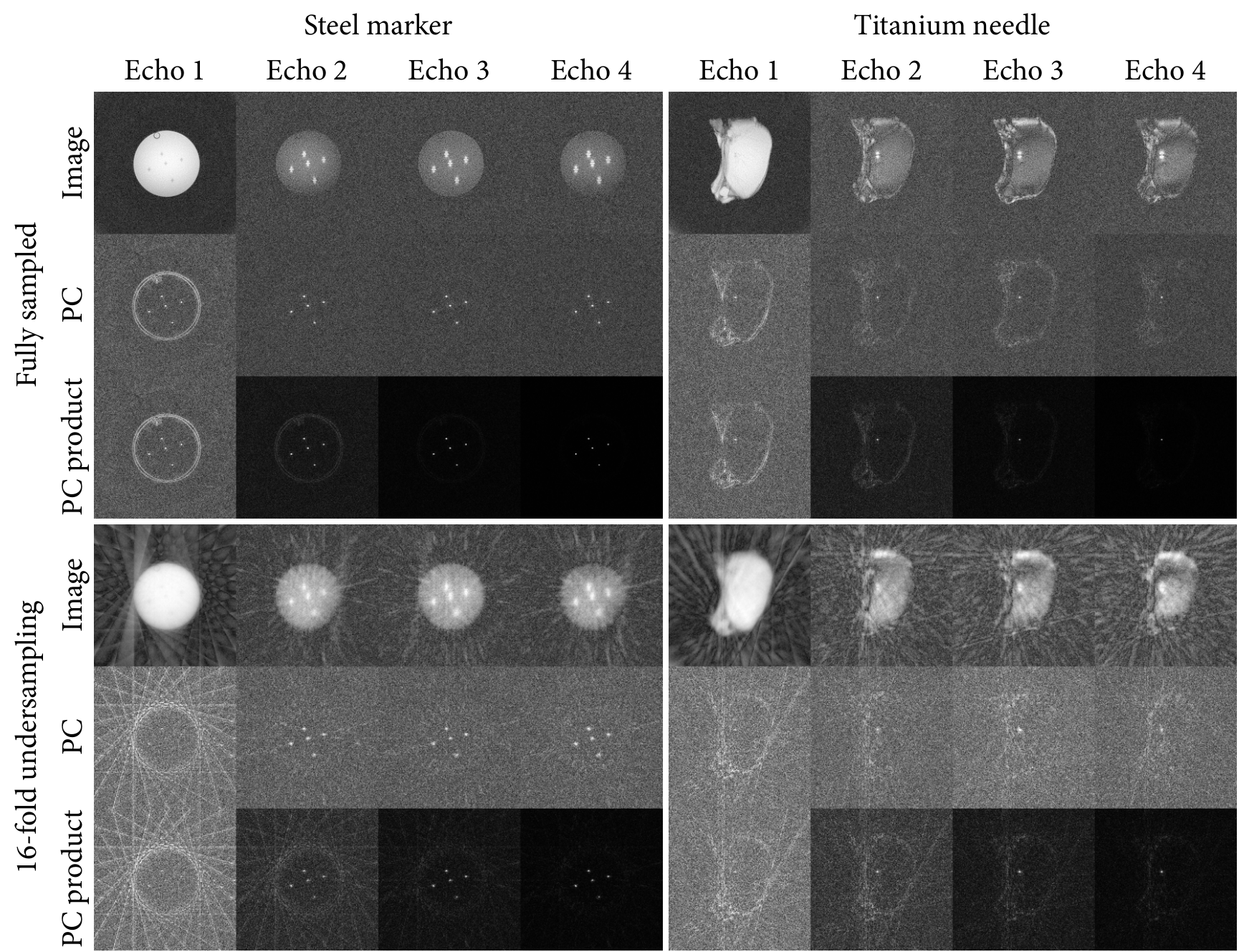}

	\caption{MRI scans and Phase Correlation (PC) maps for fully sampled (top) and 16-fold radially undersampled (bottom) acquisitions with the proposed pulse sequence for the steel markers (left) and titanium needle (right). The last row of the top and bottom sections shows the product of all PC images up to the current echo.}
	\label{fig4}
\end{figure}

Figure \ref{fig4} shows the effect of 16-fold undersampling on PC template matching using the proposed pulse sequence for the steel markers and the needle. In the echoes with white marker contrast we observed a reduction of streaking artifacts in the PC images, which suggests that the white marker contrast around devices is more specific than anatomical contrast. Most of the aliasing artifacts due to undersampling were effectively removed by taking the product of the PC images. Even at 16-fold undersampling the devices were still well-defined in the product of the PC matching on all echoes, with only minor blurring of the correlation peak.

\begin{figure}
	\centering
	
	\includegraphics[width=8cm]{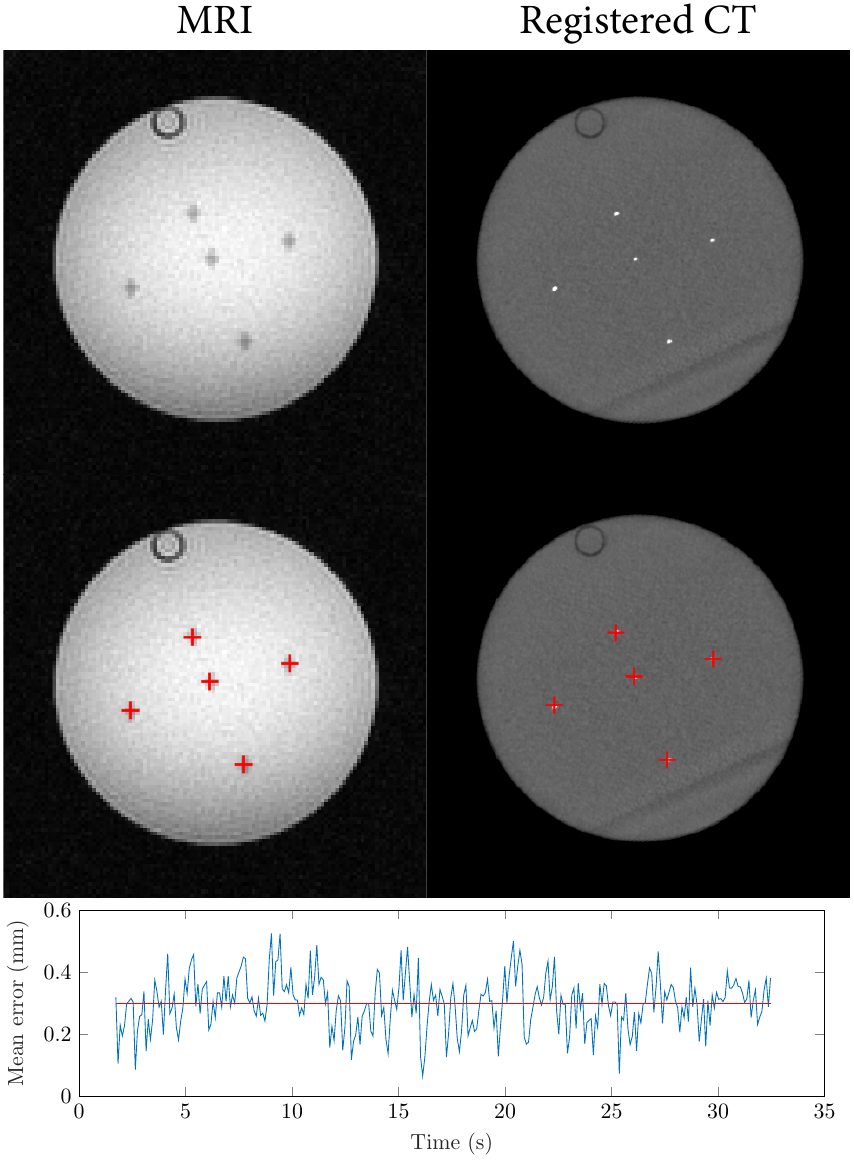}

	\caption{Steel markers in an agarose phantom localized on MRI (left) and CT (right) scans. The tracked positions of the markers at the end of the MRI tracking sequence are shown as red crosshairs on both the MRI and registered CT images in the second row. The marker positions on CT are visible as hyperintensities. The bottom graph shows the mean error of the MRI positions with respect to the CT positions over the entire tracked sequence. The mean error over all time points is shown as a red line and was 0.30 mm.}
	\label{fig5}
\end{figure}

Figure \ref{fig5} shows the localized markers (red) in the stationary steel marker phantom overlaid on the anatomical contrast for the last undersampled frame in our proposed tracking sequence, alongside the registered CT scan, on which the markers are visible as hyperintensities . The mean pairwise distance between the CT locations and the MRI locations over the entire scan sequence is shown in the bottom panel of Figure \ref{fig5}. The average distance over the entire sequence was 0.30 mm. This shows that the MR-based tracking was accurate and stable over time with respect to the configuration found on CT.

\subsection{Moving devices}

\begin{figure}
	\centering
	
	\includegraphics[width=\textwidth]{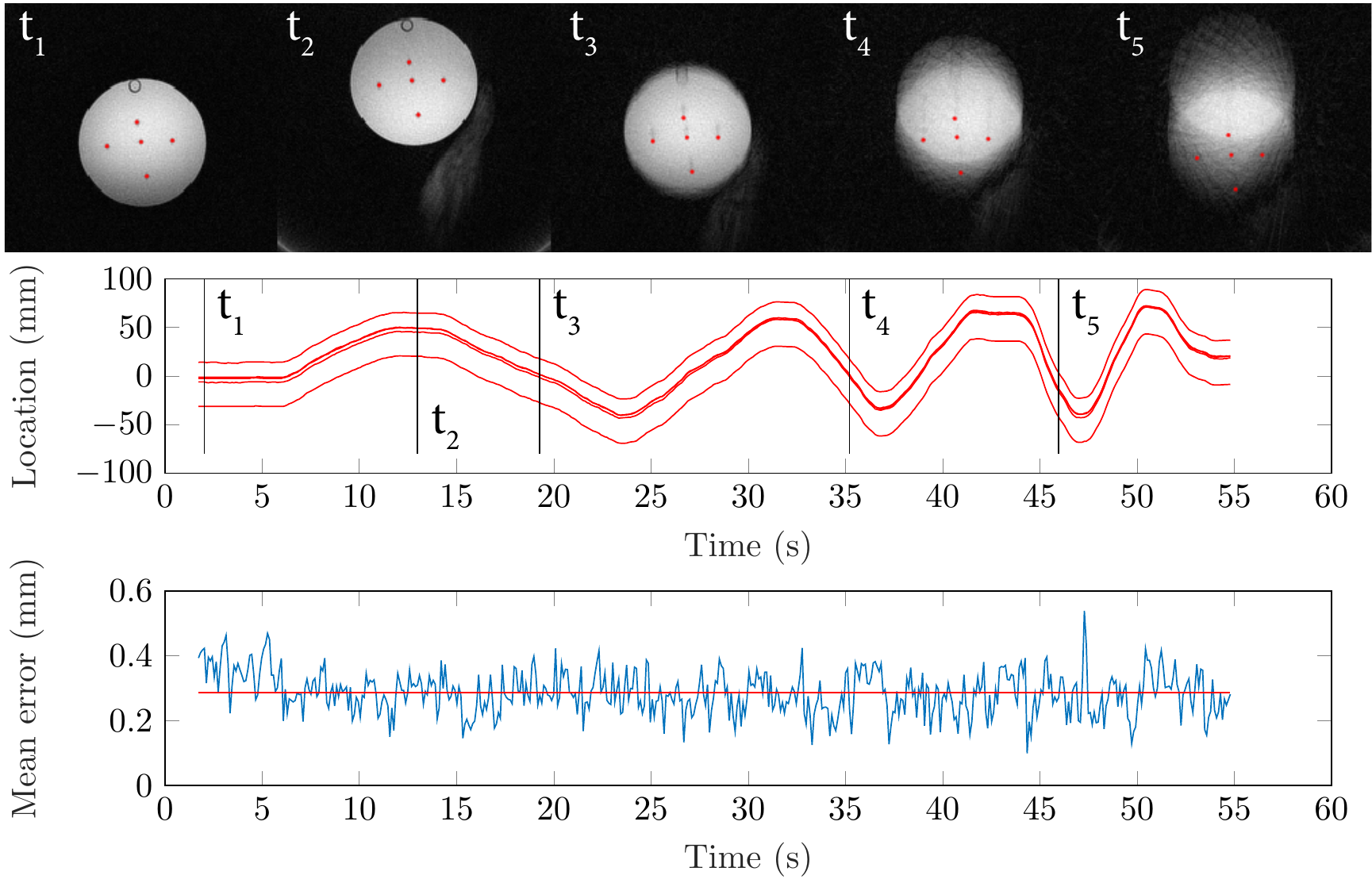}

	\caption{Tracking results for a dynamic sequence where the steel marker phantom was moving in a linear fashion along the bore of the scanner (feet-head direction). Anatomical images with the marker locations superimposed (red points) are shown for five frames (top). The middle graph shows the tracked vertical (feet-head) positions of all 5 markers over time. The bottom graph shows the mean error of the marker locations with respect to registered CT locations. The mean error over all time points is shown as a red line and was 0.29 mm. The full sequence is shown in Video 1.}
	\label{fig6}
\end{figure}

Figure \ref{fig6} shows the results of the proposed tracking method applied to a dynamic scan sequence where the steel marker phantom was manually moved in a linear fashion along the bore of the scanner (feet-head [FH] direction) at varying speeds. A video of this sequence is shown in Video 1. The FH-positions of the 5 markers all followed the motion pattern of the phantom very consistently, even at the highest speed of over 4 cm/second. However, at these unrealistically high speeds it does become apparent that the sliding window reconstruction of the anatomical image has a lower temporal resolution, which resulted in a smeared image that lags behind the tracked marker positions.

The mean error of the marker locations compared with the registered CT locations was 0.29 mm on average. The motion of the phantom did not appear to influence this error. This indicates that the configuration of the tracked markers was consistent with the configuration of the markers as found on CT, even during fast motion.

Videos 2 and 3 show the results of two additional motion sequences for the steel marker phantom are available. Video 2 shows mostly rotational motion. Video 3 shows both translational as well as rotational movement in both directions and with fast and abrupt changes in velocity. In this sequence, the tracking algorithm lost the positions of the markers at one point in time, but restored automatically after just a few frames.

\begin{figure}
	\centering
	
	\includegraphics[width=\textwidth]{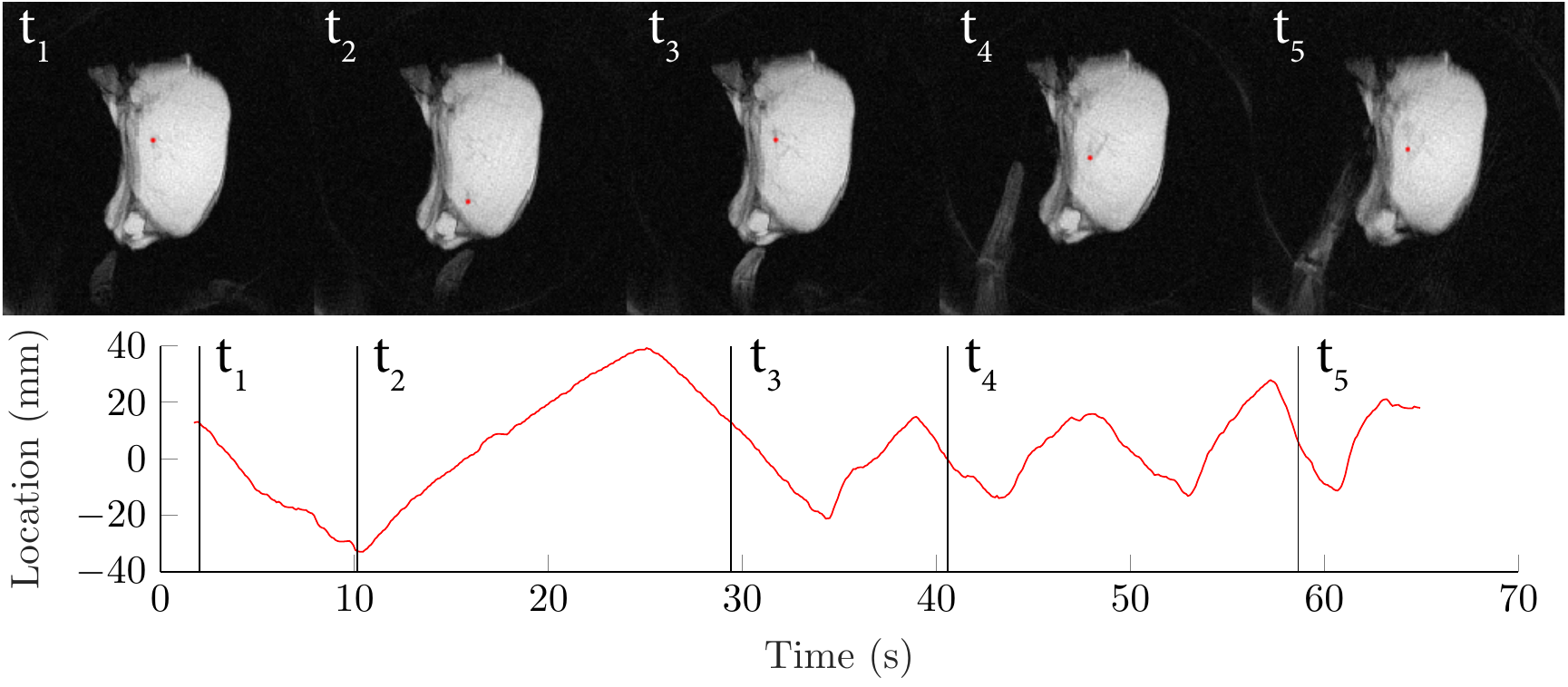}

	\caption{Tracking results for a dynamic sequence where a titanium needle was inserted into ex vivo porcine tissue along the bore of the scanner (feet-head direction). Anatomical images with the needle tip location superimposed (red points) are shown for five frames (top). The bottom graph shows the tracked vertical (feet-head) positions of the needle tip over time. The full sequence is shown in Video 4.}
	\label{fig7}
\end{figure}

Figure \ref{fig7} shows the results of the tracking method in locating the tip of a needle in ex vivo porcine tissue. The needle was being inserted and retracted along the bore of the scanner with varying speeds. The full tracked sequence is shown in Video 4. The tracked position of the needle tip was stable and smoothly moving throughout the tracking sequence, consistent with linear insertions and retractions of the needle. A sequence showing insertion of the needle at an angle of 45 degrees relative to $B_0$ is shown in Video 5. 

\begin{figure}
	\centering
	
	\includegraphics{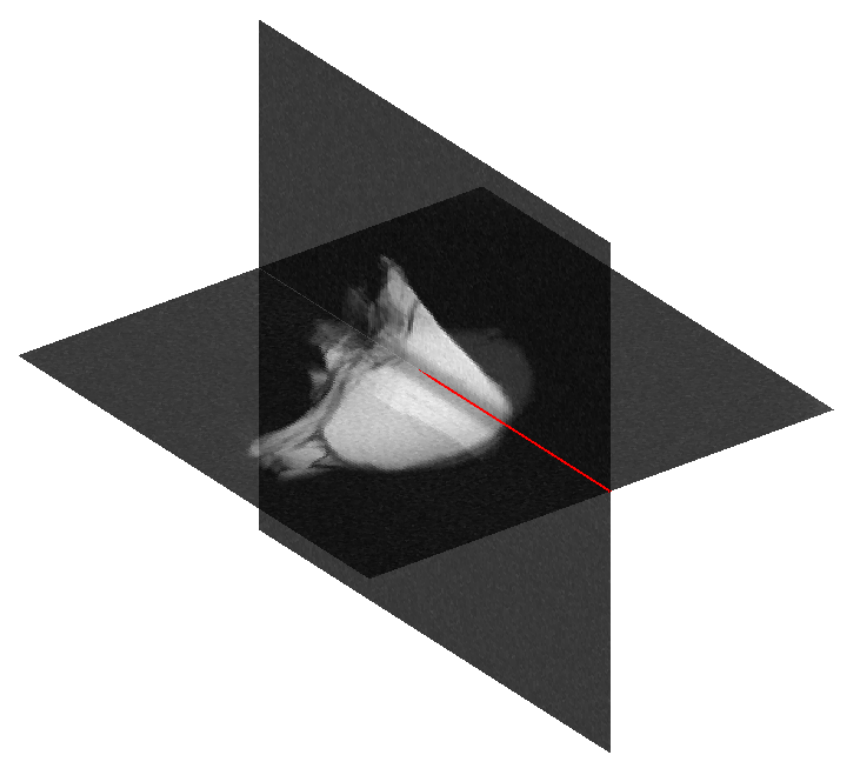}

	\caption{3D Visualization of a needle (red) in ex vivo porcine tissue that is being tracked using the proposed dual plane approach. The anatomical images are slightly transparent to allow a volumetric view of the entire field of view. The full sequence is shown in Video 6.}
	\label{fig8}
\end{figure}

Figure \ref{fig8} shows one frame of the dual plane tracking approach with a 3D visualization of the needle during insertion and retraction. The full movie is shown in Video 6. A darker band is visible in the anatomical images where the two planes overlap, which is caused by increased saturation due to overlapping RF excitations. Nonetheless, the anatomical structures are still clearly visible.

\section{Discussion}

In this study, we have demonstrated SMART tracking, a passive tracking framework that allows real-time tracking of metal devices in 2D MRI at a rate of 10 updates per second, while also providing a sliding window anatomical image with a temporal resolution of 1.6 seconds. Because both the device localization and the anatomical image were simultaneously acquired with the same pulse sequence, the tracked positions of the devices were intrinsically registered to the anatomical image. This was accomplished by combination and extension of previously proposed methodologies for positive contrast, acceleration, device localization, simulation, and tracking. The combination of these methods provides advantages over the individual methods, by increasing temporal resolution, robustness and flexibility.

The mean localization error for the stationary phantom was 0.30 mm, which was near the theoretical limit of a system that localizes on the MRI resolution of 1.2 mm. This limit may be overcome by performing PC template matching at sub-voxel resolutions \cite{alba_phase_2015}, which may further improve the accuracy of our method. We observed only minor differences in localization errors between stationary and moving experiments, which suggests that the method is robust with respect to motion of devices. The tracked positions appeared to be robust at velocities much higher than those that would be expected in interventional applications. 

In the first echo of the acquisition, SMART tracking provides a $T_1$-weighted sliding window anatomical image. The short echo time and high readout bandwidth minimized the off-resonance artifacts around the devices and provided a relatively undistorted view of the anatomy around the device that could be used to identify the target region during MR-guided interventions. The temporal resolution of the anatomical image was 1.6 seconds, which does mean that fast moving anatomies, such as the heart, will appear blurry. However, we expect good image quality in anatomical regions where motion would be slow, such as the pelvis and the brain, for example in biopsies of the prostate or uterus, and in placement of deep brain stimulators. Application of Compressed Sensing reconstruction to the undersampled radial data may help improve the temporal resolution to provide better image quality in faster moving anatomies.

In this proof of concept study we designed the pulse sequence such that the device tracking was robust in the experiments demonstrated. However, the SMART tracking framework allows various modifications that could accelerate the method, improve SNR, or improve tracking quality. For example, the number of echoes could be reduced to accelerate the acquisition, providing higher temporal resolutions, or allowing lower readout bandwidth for higher SNR. Additionally, the amount of dephasing in each echo can be varied over the echoes, which may provide better white marker contrast, especially for larger devices. Interestingly, using scanners with higher field strengths would allow smaller devices to be tracked, because magnetic susceptibility artifacts scale with the main magnetic field strength \cite{ludeke_susceptibility_1985}. Furthermore, the methods we described are independent of the sampling scheme and could be applied to conventional Cartesian acquisitions, or other non-Cartesian acquisitions, such as spiral sampling. Investigation into the impact of these choices was outside the scope of this study, but might improve the tracking methodology, especially when optimized for the requirements of a specific interventional application.

The use of simulation and template matching for device localization offers flexibility with regard to the device being tracked. The main requirement is that the artifacts around the device can be accurately simulated in a single template. This does assume that the orientation of the device is constant, for example during a linear insertion of a needle, or that the artifacts are independent of orientation, for example in a spherical marker. In cases where the orientation of the device is not constant, for example when the tip of a needle deflects, a library of templates for multiple orientations of the device could be simulated \cite{karimi_position_2016,wachowicz_characterization_2006,zijlstra_fast_2017}. 

Finally we have shown that SMART tracking is directly applicable to dual plane tracking, where the slice orientation of every undersampled frame is alternated between the two planes. This lowered the anatomical framerate by a factor 2, but enabled the method to track the position of the device in all three dimensions. In a system integrated with the scanner software, this dual plane approach would allow automatic slice positioning to keep the scanning planes centered on the device. 

\subsection{Limitations and future work}

In this study we have demonstrated the basic principles of SMART tracking in phantom experiments. While the results are promising, it is important to note the limitations of this study with respect to tracking devices in clinical conditions, and the steps necessary for clinical implementation and validation.

First, the experiments in this study provide limited insight into the accuracy of the sliding window anatomical images. Our experiments involved fairly homogeneous phantoms with rigid motion, while in a clinical application tissues will be more heterogeneous, and deformable motion of tissues is likely to occur. Given the relatively low framerate of the anatomical image, it should be investigated under what conditions a target region, such as a tumor, can be identified during an intervention.

Second, the imaging plane in our experiments was static. Therefore, a relatively thick slice of 15 mm was needed to keep the device in-plane, which limits the localization accuracy in the slice direction. In clinical applications, a device may move out of the imaging plane, which requires the ability to dynamically update the plane position, either manually by an operator or automatically with a dual plane acquisition. Furthermore, if the operator has the ability to switch between two orthogonal plane orientations, the device can also be accurately localized in the slice direction when necessary.

Finally, magnetic field inhomogeneity effects can be more severe in vivo than in phantom experiments, for example near the lungs. In this study, we observed limited effects of field inhomogeneity in the white marker images, which did not appear to influence the PC images (Figure \ref{fig3}). Whether the field inhomogeneities that are present in vivo would significantly influence the proposed tracking methodology can only be determined by in vivo experiments.

A challenge for clinical validation is the implementation of SMART tracking in a real-time setting. Real-time access to raw k-space data is necessary to perform our image reconstruction and tracking. Furthermore, the operator needs to be given the ability to reposition and reorient the scan in real-time to keep the device visible. And finally, the anatomical images and device location should be presented to the clinician performing the intervention with minimal latency. While these technical challenges are not necessarily hard to solve, it does require time and cooperation from the scanner vendor. 

\section{Conclusion}

SMART tracking shares some of the advantages that are traditionally only available with active tracking methods: accurate tracking of devices at high framerates, inclusion of real-time anatomical scanning, and, with the dual plane approach, automatic slice positioning. Yet, the proposed method does not require specialized hardware. Instead, any rigid metal device that is safe for MRI and that causes appreciable magnetic field distortions can be tracked using this method. While more experiments are required to definitively prove the robustness of the method in clinical applications, the results in this study show promise for a flexible, low-cost approach to MR-guided interventions.

\acknowledgments 

This work is part of the research programme Applied and Engineering Sciences (TTW) with project number 10712 which is (partly) financed by the Netherlands Organization for Scientific Research (NWO).

\section*{Appendix A. Implementation details}

\subsection*{Non-Cartesian simulations with the white marker phenomenon}
The simulations in this study were performed with the FORECAST method \cite{zijlstra_fast_2017}, which we extended to allow simulation of the white marker phenomenon and non-Cartesian sampling trajectories. Because the white marker phenomenon is similar to a phase encoding gradient, the effect of this gradient can be included in the signal equation used in FORECAST as a single encoding step in the $z$ direction:

\begin{equation}
s(k_x,k_y) = \sum_{z} \sum_{y} \sum_{x}  \rho(x,y,z,t') e^{i 2 \pi \Delta B_0 (x,y,z) t' + k_z z} e^{-t' / T_2(x,y,z)} e^{-i 2 \pi (k_x x + k_y y)}
\end{equation}

To allow non-uniform sampling the method was modified to use a non-uniform fast Fourier transform (NUFFT) \cite{fessler_nonuniform_2003} for each unique time point $t'$ to evaluate the Fourier encoding for all frequencies ($k_x$, $k_y$) sampled at time point $t'$. This approach is slightly less efficient than the original approach for Cartesian trajectories, because the readout and phase encoding cannot in general be separated in a non-Cartesian acquisition. Nevertheless, the speedup relative to Bloch simulation is still expected to be in the order of the number of repetitions of the pulse sequence.

\subsection*{Device tracking}
The device tracking in this study was performed with a Kalman filter \cite{kalman_new_1960} with a linear motion model that includes the position and velocity of the device:

\begin{equation}
x_{i+1} = x_i + \mathit{dt} v_i + N_x
\end{equation}
\begin{equation}
v_{i+1} = v_i + N_v
\end{equation}

Here, $x$ is the 2D position of the device, $v$ is the velocity vector, $dt$ is the discrete time step of the model, and $N_x$ and $N_v$ describe normally distributed process noise (i.e. how fast the values are allowed to change). The measurement with which the model was updated was a single measured position $z$ per frame: $z_i=x_i+N_z$, where $N_z$ describes normally distributed measurement noise. The process and measurement noise covariance were experimentally determined. Process noise was set to a standard deviation of $1 \cdot dt$ voxels for the position, and $10 \cdot dt$ voxels per second for the velocity. Measurement noise was set to a standard deviation of 0.5 voxel.

For each frame a position measurement was extracted from the PC image by finding a location that optimizes both the PC intensity and the distance to the previously predicted location of the device. Candidates for these locations were required to be a local maximum in a $7 \times 7$ region in the PC image. In the case of tracking multiple identical devices, we used the Hungarian Method \cite{kuhn_hungarian_1955} to optimally assign a candidate position to each device.

\bibliography{report} 
\bibliographystyle{spiebib} 

\section*{Videos}

\begin{itemize}
\item \href{https://www.isi.uu.nl/People/Frank/Thesis/movie1.html}{Video 1}: SMART tracking applied to the steel marker phantom moved in a linear fashion along the bore (feet-head direction) with varying speeds.

\item \href{https://www.isi.uu.nl/People/Frank/Thesis/movie2.html}{Video 2}: SMART tracking applied to the steel marker phantom rotated with varying speeds.

\item \href{https://www.isi.uu.nl/People/Frank/Thesis/movie3.html}{Video 3}: SMART tracking applied to the steel marker phantom moved and rotated with high speeds.

\item \href{https://www.isi.uu.nl/People/Frank/Thesis/movie4.html}{Video 4}: SMART tracking applied to insertion and retraction of a titanium needle in porcine tissue with the needle parallel with $B_0$.

\item \href{https://www.isi.uu.nl/People/Frank/Thesis/movie5.html}{Video 5}: SMART tracking applied to insertion and retraction of a titanium needle in porcine tissue with the needle at an angle of approximately 45 degrees with $B_0$.

\item \href{https://www.isi.uu.nl/People/Frank/Thesis/movie6.html}{Video 6}: Dual plane SMART tracking applied to insertion and retraction of a titanium needle in porcine tissue with the needle parallel with $B_0$.
\end{itemize}

\end{document}